\documentclass[rmp,10pt,showpacs,reqno]{revtex4}

\usepackage{amsmath,amscd,amsfonts,hyperref}

\usepackage[final]{graphicx}

\DeclareGraphicsExtensions{.eps,.pict}%
\newcommand{\Rset}{\mathbb{R}}
\newcommand{\Nset}{\mathbb{N}}
\newcommand{\m}{\mathcal{M}}
\newcommand{\V}{\mathcal{V}}

\newcommand{\tot}{\mathcal{T}}

\newcommand{\tom}{T_{p_0}\m}

\newcommand{\wg}{\widehat{\gamma}}
\newcommand{\wc}{\widehat{\mathcal{C}}}
\newcommand{\wl}{\widehat{\mathcal{L}}}

\newtheorem{definition}{Definition}

\begin{document}

\title{Time `betwins'}

\author{Thierry GRANDOU}
\email{Thierry.Grandou@inln.cnrs.fr}

\author{Jacques L. RUBIN}
\email{Jacques.Rubin@inln.cnrs.fr}

\affiliation{Institut du Non-Lin{\'e}aire de Nice,\\
U.M.R. 6618, C.N.R.S. - Universit{\'e} de Nice - 
Sophia-Antipolis,\\
1361 route des Lucioles, 06560 Valbonne, France}



\begin{abstract} 
\textbf{Abstract:} 
Discussions on the Langevin Twins `paradox'
are most often based on a ``triangular" diagram which outlines the
twins spacetime travels. It won't be our way, avoiding what we
think to be a problem at the basis of numerous
controversies. Our approach relies on
a fundamentally different Equivalence Principle, namely the
so-called \textit{``Punctual Equivalence Principle"}
\cite{ghinsbuden2001}, from which we think that a very
conformal aspect proceeds.
\par
We present a resolution of this paradox in the framework of
the so-called \textit{``scale gravity"}.  This resolution
hinges on a clear determinism of the Twins proper times, in
some definite situations, and a fundamental
under-determinism in some other particular ones, the
physical discrimination of which being achieved out of a
precise mathematical description  of  conformal
geometry.\par Moreover, we find that the time discrepancy
between the twins, could somehow be at the root
expression of the second fundamental law of thermodynamics.\par
\bigskip
\parbox{13cm}{\textbf{Work presented at the ``1rst International Conference on
the Ontology of Spacetime",\\ 11--14 May 2004, University of Concordia, Montr{\'e}al (QC),
Canada. \\ \\
{\large PREPRINT INLN \#2004/11\/}}}
\end{abstract}

\maketitle


\section{Triangular diagrams}
Very often, discussions on the Langevin twins paradox are
based on a ``triangular" diagram, and
we would like, first, just
to recall a little problem of distance measurement,
evaluation and/or conception in spacetime,
which appeared in one of the lightest ``triangle"
approach given by H. Bondi \cite{bondidisc57} and 
in fact historically due to
Lord Halsbury (see H. Dingle's work in \cite{changstud93}) and
involving three unaccelerated observers.
\par
Bondi considered  diagrams (refered as the Fig. 2 and
Fig. 3 in his 1957 paper) that can be assembled in a
unique diagram as follows (Figure 1.) with the so-called
apex at the crossing point Y:
\begin{center}
\noindent\fbox{\includegraphics[scale=.95]{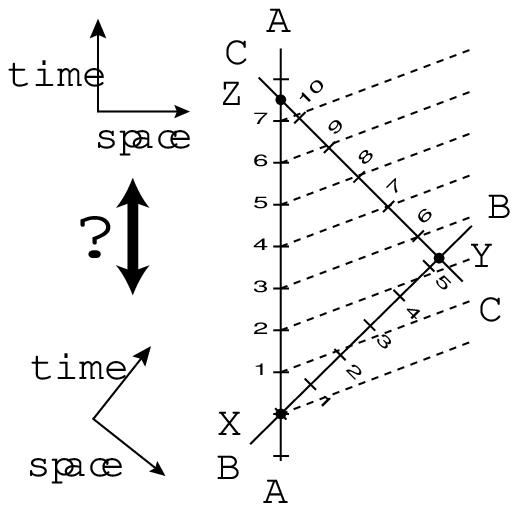}
Figure 1.}
\hskip1cm
\fbox{\includegraphics[scale=.95]{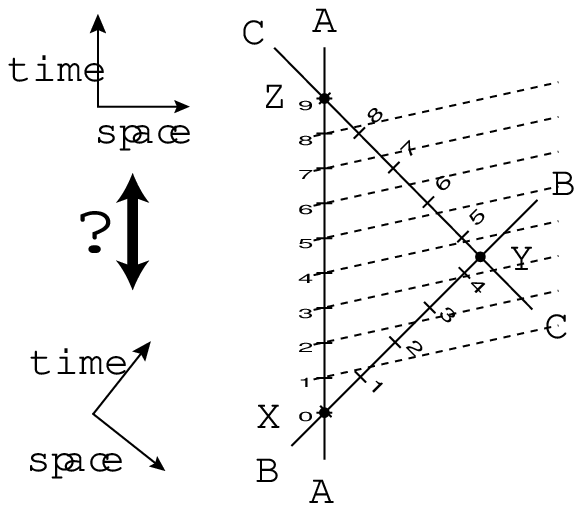}
Figure 2.}
\end{center}
The ticks of all the clocks are represented by numbered
marks on the three straight lines AA, BB and CC, and at
each tick a ligth ray is emitted isotropically and
is represented by a dashed line. In this diagram the
inertial on-earth twin observer represents himself
on the line AA. The traveller twin observer
moves along the line BB from point X to point Y where
he transmits his proper time to a third crossing twin
traveller coming back towards the on-earth twin 
along the line CC from point Y to point Z. The
speeds of each travelling twin along lines BB and CC
are assumed to be obviously non-vanishing, the same
and constant. Also, and that is the problem we focus
on, Bondi implicitly assumed that the lenghts between
consecutive tick marks on the lines are all the
same. Clearly from the latter figure, at point X,
then each twin, comparing ticks intervals of his
clock to the other twin one by measuring intervals
between light rays arrivals on his own line, would
get longer observed time intervals due to the
well-known spacetime parallax, i.e. the time
dilatation which can be obtained applying Lorentz
transformations. Hence each twin thinks the other has
a slower time. That is a clock ``paradox" which is not
a logical one, but based on an incompletness of the
special or general relativities to determine, because
of reciprocal viewpoints in formulations of each
of these relativities, who is the oldest or younger
twin. Also, how must be the spacetime axis orientation
choice, as indicated in the figures, i.e. the so-called 
space-time spliting ? \par\medskip
But, the same can be obtained for intervals between
marks varying in a particular range in such a way
that the equality of length between marks
on each line, or passing from one line to another
one, would be broken. Hence we could obtain a kind
of figure as above (Figure 2.) with again a parallax (a time
dilatation) at point X, but 
an equal time for the twins
when going on from X to Z ! And then there is no
twins paradox. That is no more than the classical acoustic Doppler
effect !\par What we want to point out in this
example, is the following question:\par\medskip 
$\bullet$ How to physically compare distances on different paths~?\par\medskip
By the twins
experiments ! Indeed, we have no ubiquity property 
which would allow us to be ``simultaneously" on two 
different paths to make comparisons, superpositions
of extended objects ! Also, we have a very different viewpoint
about the concept of ``absolute simultaneaous events", and
we agree about the conventionality of simultaneity
as T. A. Debs and M. L. G. Redhead described it
\cite{petkov87,debsredhead96}.\par\medskip
$\bullet$ Is acceleration a	fundamental  tool to solve the twin
paradox and/or to reply to the previous question ?\par\medskip
We know that A.~Einstein (1918; See p. 43, \S 3 and \S 4 in 
\cite{prokhovnikbook67}), invoked  1) the
space traveller twin acceleration at the ``apex" of
the triangular diagram  where he comes back, 2)
together with the ``elevator metaphor", i.e. the
so-called Einstein equivalence principle (see
a complete review of this principle in \cite{ghinsbuden2001}), 
to justify the time dilatation of the travelling twin proper time from
the equivalence of the travelling twin acceleration with a
gravitational field effect in a \textit{``static"\/} situation.\par
Then to reply to this question,  we must deal firstly with a general 
relativity description without
accelerations. It would imply a kind of fundamental  change
of terminology as well as in the reasoning.
\par\medskip   
Let us remark that the Einstein's explanation contradicts the
given one proposed by H. Bondi in his 1957 paper, since he
showed with the ``three brothers paradox" that
acceleration is not necessary to explain the latter
time asymetry. Hence, this paradox would appear at
the special relativity ground level and would have to
be solved, at this stage, without the use of acceleration
but with a general relativity description.\par 
That was the H. Bondi standpoint, and solving a special
relativity effect with help of the general relativity would
be a first contradiction only if and only if paths
considerations are admitted in the framework of
special relativity, what we reject as precised
further and we would have to accept 
the Einstein arguments.\par \medskip
But in fact, there is another contradiction
(!) in the Bondi solution quoting the deep 
H. Dingle remark (in spite of a large amount of
contradictions in Dingle's works !) based on the Einstein
foundations of the special relativity 
\cite{dingle57} (See also quoted by G.
Builder in \cite{builder58}):
\begin{quote}
\emph{``It should be obvious that \emph{[if]\/} there is an
absolute effect which is a function of velocity then
velocity must be absolute. No manipulation of
formulae or devising of ingenious experiments can
alter that simple fact."\/}
\end{quote}
But the problem coming after, was to expand the
latter Dingle remark to acceleration, \textbf{and 
thus to the Einstein arguments}. Mendel Sachs
did it, and in fact at any order of the ``proper
time derivation" (pp.94-95 in \cite{sachs179}):
\begin{quote}
\emph{``Einstein tried to resolve this paradox by
taking into account the periods of non-uniform
motion during the round trip journey.\dots\,\, \dots
it implies that while \emph{[velocity]} is a
relative dynamical variable,
\emph{[acceleration]} is an absolute dynamical
variable (since it acts as the cause of an
absolute physical effect). But this is not true,
according to Einstein's theory of relativity ! If
the spatial and temporal coordinates are all
relative to the reference frame in which they are
expressed (in contrast with the ``absolute"
temporal coordinate of classical physics), then
\emph{[any]} order derivative of any of these
coordinates with respect to any other must also
be relative.\dots \emph{[That is]}, 
acceleration is as relative as velocity is."\/}
\end{quote}
This has been clearly demonstrated by Unruh \cite{unruham81} in the framework of the twin paradox and in considering the acceleration at the apex of a
triangular diagram. He shown that the latter acceleration is  absolutely correlated with a sudden kinematical acceleration of the on-earth twin seen by the
travelling twin. Hence, neither the Bondi nor the Einstein arguments can be used, and in fact Einstein would contradict himself because of the elevator
metaphor and the relativity principle~!\par\medskip
Moreover from the Einstein elevator metaphor, we can't be
able to distinguish in an acceleration the part due to
gravitation from the one due to kinematics ($\ast$). Hence acceleration
is not the main characteristic of gravitation. Roughly
speaking, gravitation may be the efficient cause to 
phenomena correlated to relative accelerations,  not the
converse.\par
But then from ($\ast$), what is gravitation if
acceleration of gravitation is not the main
characteristic~? 
\par\medskip
On the other hand, we consider that paths
are truly and only described in general relativity and we make a difference
between time and duration following the H. Bergson philosophy. And if there
is an absolute duration, the latter can't be associated to
gravitation which must also be relative.\par\bigskip
It seems that the Twins pseudo-paradox is only a problem coming from special relativity and it wouldn't have to be solved in a general relativity framework. 
Indeed, it appears that it is only a pure chronogeometrical effect for which general relativity is of no relevance (at least in the standart exemple).
The Minkowski spacetime would be the coherent and suitable framework to compute the proper time on every worldline, right or not (otherwise, it would be
equivalent to forbid  computation of lengthes of curves in the Euclidean geometry !).
In this latter framework the acceleration wouldn't be relative since only frames in a relative uniform unaccelerated motion are equivalent.
In fact, we can consider that in each frames, relative acceleration can be measured and defined. The acceleration doesn't appear to state an equivalence
between frames, thought it would be the case in the Einstein relativity rather than in the Galilean relativity. That is the meaning of the Einstein elevator
metaphor. Moreover, the heart problem is that the proper time of the special relativity can only be evoked for parallax effects, i.e. effects coming from
comparison of vectors and not curves. If the Twins pseudo-paradox would come from a parallax effect, then it would have both experimental and theoretical
reciprocity, expressed from the reversibility of the Lorentz transformations applied to vectors (in a tangent space) of the special relativity. Then the theory
would be complet.\par
In fact special relativity involves to consider only vectors with same base point and not curves. If this would be possible, we would be able
to deduce lenghts of curves from their projections on planes (the tangent spaces) with their origins being the measurement events. It would be as if we would 
deduce lengths of curves from their projections ! We would make a lot of errors in our lentghs evaluation, unless to know the projections. It would be as to
deduce the length of a road from a plane picture of the road. In fact, this latter metaphor gives the way  to solve the Twins pseudo-paradox : how to
know the projections. Hence, in order to make the lenght of curves evaluation, we need to move along the worldlines and to be able to deduce how the
displacement and the projections all along are. It is a kind of projective geometry effect. Then considering curves and not vectors, we need to be in the
general relativity framework even if acceleration are not considered.
\section{Observers tied with their spacetime
environment: a punctual principle  of equivalence}
Hence, an observer would be tied with a spacetime
environment. Which one and what does it 
mean exactly ? 
What is at the core of a \textit{``state of motion"\/} concept ? How to evaluate
distances ? \par
The coordinate maps, from the spacetime to
$\Rset$ (vector) spaces are only defined on those
subspaces of points at which serial discret physical
measurements (labeling, counting, coding, etc\dots)
are performed. Only distances could be defined on
those subspaces. These maps can be defined on some
neighborhoods and then on open subsets. Hence, most
of the ``time" we can only deduce a local metric
at a point, not a field of metric in a spacetime, meaning that we have only values  of \textit{germs\/} of metric fields at a point, i.e. \textit{0-jets\/}, not
defined on $\m$ in full generality. In fact, more generally, we have a metric attached to each point of
a line, a trajectory of a spacetime ship, and this
metric would be built out of a local moving frame 
attached to material probes (i.e. objects). 
\par The tangent spaces $\tom$ of moving frames, will be viewed
as the spaces for comparison between given objects and rulers, callipers and watches, ``unfolded" on $\m$. These two
separate categories, namely the rulers and those which are ruled, can be viewed as \textit{``twins"\/} categories, and each of these objects are
associated to different \textit{jets\/} of sections of $T\m$ defined on each given open set $U\subset\m$. Hence, each measurement necessarily involves twins objects
each one associated to a unique ascription of a velocity to a point in $\m$. We think that this relation between a velocity and a point, is thus what is
defining a classical object.
\par\medskip
Then the punctual equivalence principle, we think being compatible with the previous remarks, is
the following. Let us assume the  4D-ocean
$\m$  to be a manifold of class
$C^2$, of dimension 4, locally arc-connected and paracompact (i.e. Hausdorff and union of a countable set of compacts).
Let
$p_0$ be a particular point in
$\m$, $U(p_0)$ an open neighborhood of $p_0$ in $\m$, and $\tom$ its
tangent space. The \emph{``punctual"\/} 
principle of equivalence we use,
states it exists a local proper diffeomorphism
$\varphi_0$ (assumed to be of class $C^2$), that we call the
equivalence map, attached to
$p_0$ putting in a one-to-one correspondence the points
$p\in U(p_0)$ with some vectors $\xi\in\tom$ in an open neighborhood of
the origin of $\tom$:
\[
\varphi_0:p\in
U(p_0)\subset\m\longrightarrow\xi\in\widehat{U}_0\equiv\varphi_0(U(p_0))\subset\tom\,,\qquad
\varphi_0(p_0)=0\,.
\]
In fact, we consider that only shapes are kept,
i.e. projectable on ``board", i.e. on $\tom$, and that we can't
evaluate distance ratios passing from $\m$ to
$\tom$ (somehow the standarts can't be compared beacause  of 
non-ubiquity).
In some way, it would be  a consequence of a kind of
time travel inability and/or time non-ubiquity, since
traveling in time and time ubiquity would be needed to
make comparison of distances, ``simultaneously" in
$\m$ and $\tom$.
\begin{center}
\noindent\fbox{\includegraphics[bb=-70 235 585
754,scale=.28]{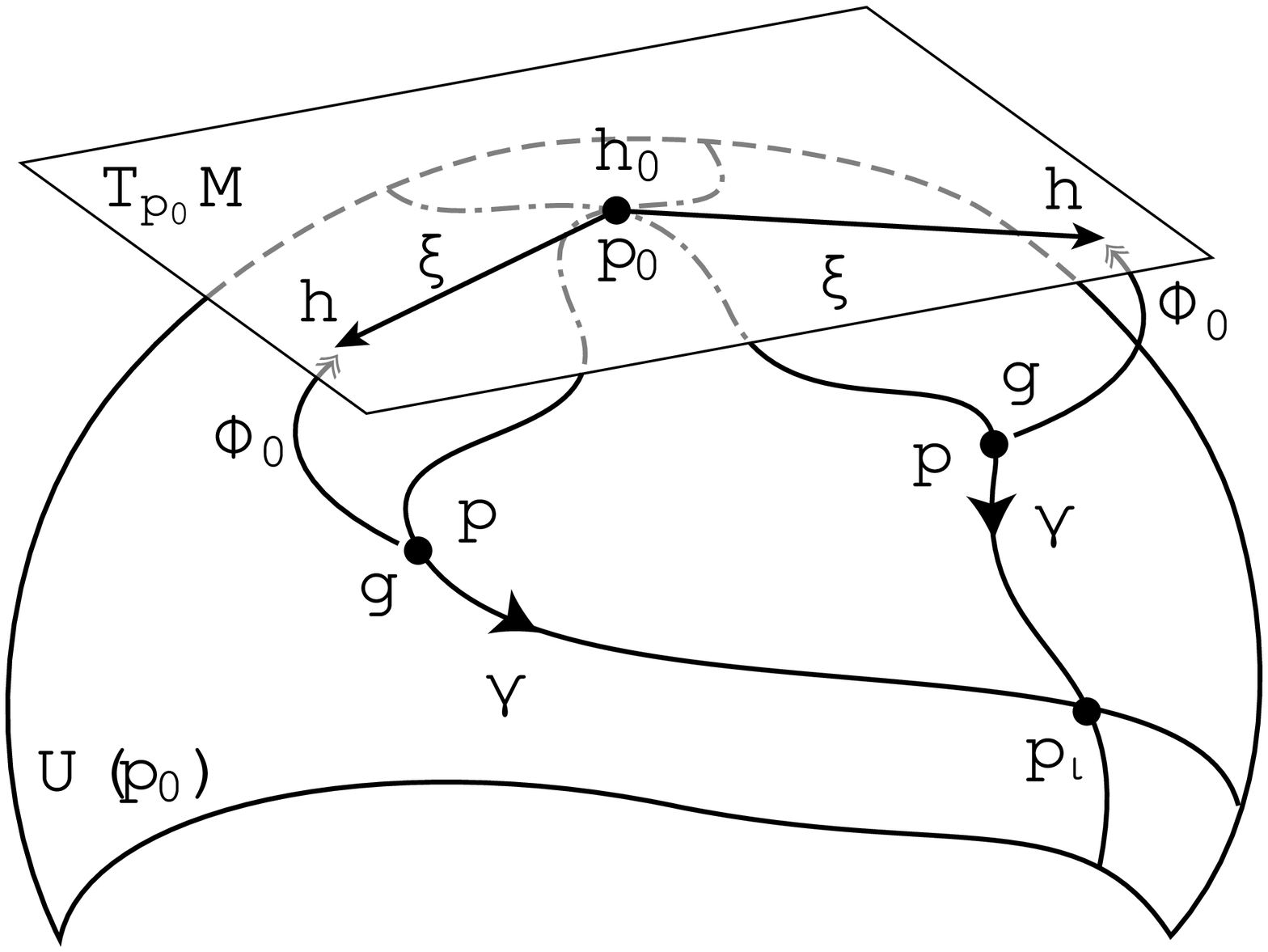}
Figure 3.}\hskip5mm
\noindent\fbox{\includegraphics[bb=-70 235 585
754,scale=.28]{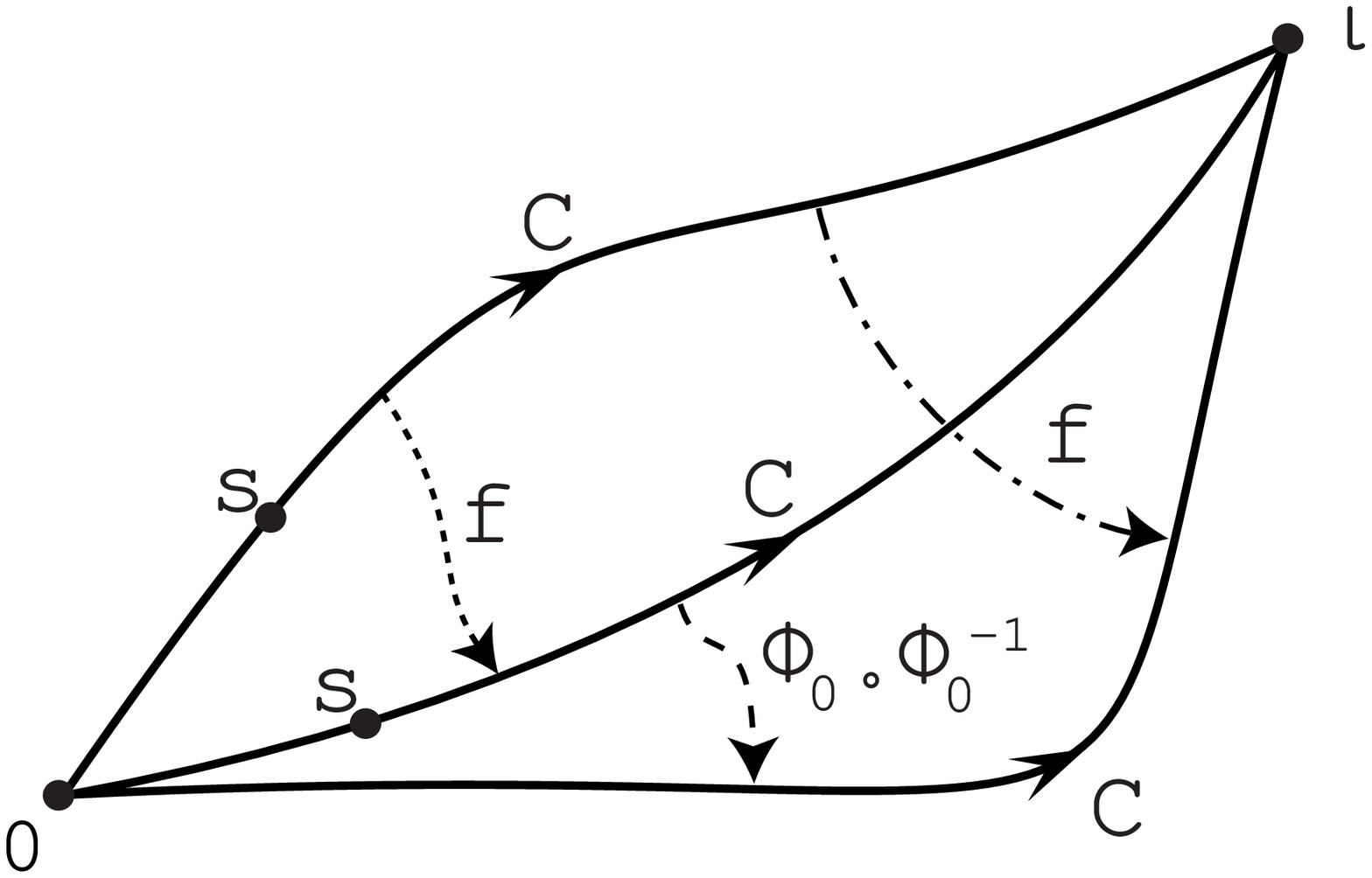}
Figure 4.}
\end{center}
Here in the figure above (Figure 3.), the point $p_0$ (other points
could be chosen) is the
crossing point of two  worldlines, which can be geodesic
or not, and two particular points $p$ and $p'$ are taken respectively on
the latter. The spacetime $\m$ is assumed to be locally
analytic, i.e. we can make local analytic charts, in
particular in a neighborhood $U(p_0)$ of $p_0$.
Consequently, $\varphi_0$ is an analytic map. From
this local  description, we give, in the
following subsection, the transformation laws
related to conformal or projective geometry and from which
special relativity can be deduced and generalized. It has to be noticed that conformal geometry involves angle  measurements \textit{via\/} stereometry
(telemetry) as W. Unruh clearly demonstrated in his approach of the Twins paradox \cite{unruham81}.\par
\medskip
An important remark has to be done about significance of ``path",
since what means ``path" for an unique object in an empty spacetime
? If two objects, we can only speaks about  a relative path and the
two paths drawn on Figure 3 (paths $\gamma$ and $\gamma'$) is a
physical non-sense, unless a recording observer at a crossing event at $p_0$ exists, and from which  the two relative paths are ``assessed".\par\medskip
We emphasize that all in our geometrical presentation is local. Then, we will only consider jets of maps defined on opens in $\m$, and never  maps on $\m$
itself, meaning that the global geometrical or topological structures of $\m$ won't be considered, but its local \textit{{\'e}tale\/} presheaf
structures only. Hence by considering $\m$, in fact we  consider  a given open subset $U$ with the same local topological properties~: Hausdorff, paracompacity,
etc\dots  and by $\m$ we have to consider the full set (which is an equivalence class) of manifolds with the same local topological structure on $U$ and
local rings of maps on $U$. In other words we would denote $[\m](U)\equiv[\m]$ instead of $\m$ to indicate the class (i.e. the germ of manifolds on $U$
endowed with the map $\varphi_{p_0}$) of
$\m$. In order to lighten the notation we only denote this class $[\m]$ by one of its representant, namely $\m$.
\section{Conformal affine transformation laws: special relativity
with acceleration}
This chapter can be skiped at a first lecture, since it deals with the demonstration to find smooth paths linked by an isometric function $f$.\par
Then, we consider a Lorentzian metric
$g$ (not a field of) at point $p\in\m$, and its corresponding
Lorentzian metric $h$ at vector $\eta$ such that $\varphi_0(p)=\eta\in\tom$,
$\varphi_0(p_0)=0$ and
$\varphi_0^{\ast}(h)=g$\,; The same for $p'$ but
with prime symbols adding up to the previous
relations, but considering only one equivalence map
$\varphi_0$ as shown on the figure above. We
denote $\tot\equiv\tom$.\par
At a first step we consider paths as $C^{2}$ not-piecewise proper
embeddings (no self intersections  contrarilly to immersions) 
and consequently without critical points 
(any inverse image of a compact is a compact), included in
$U(p_0)\subset\m$:
\[
\gamma\,:s\in[0,1]\longrightarrow p\in\mathcal{C}\equiv\mbox{Im}\gamma\subset U(p_0)\,,
\qquad
\gamma'\,:s'\in[0,1]\longrightarrow
p'\in\mathcal{C}'\equiv\mbox{Im}\gamma'\subset U(p_0)\,,
\qquad
\gamma(1)=\gamma'(1)\equiv p_\iota\,,
\]
with $U(p_0)$ simply connected and \textit{relatively compact\/}, i.e. $\bar{U}(p_0)$ is compact (it
will insure $U(p_0)$ to be bounded), and assumed simply connected. Moreover we consider in the sequel, paths with only 2
intersections in
$U(p_0)$. We denote
$\xi\equiv\wg(s)\equiv\varphi_0\circ\gamma(s)$ and
$\xi'\equiv\wg'(s')\equiv\varphi_0\circ\gamma'(s')$ the
corresponding paths on
$\tom$, with
$\wg(0)=\wg'(0)=0$,
$\wg(1)=\wg'(1)=\iota$,
$\wc=\wg([0,1])$ and $\wc'=\wg'([0,1])$. 
The conformal assumption
we make, from the identification of $T_\xi\tot$ (resp.
$T_{\xi'}\tot$) with $\tot$, and which lighten the discussions
below and thus motivating to work on $\tot$ rather than on $U(p_0)$,
states that only (\textit{``a priori"\/}) if the vector
$\xi$ is  an element
of $\wc$ (resp. $\xi'$ an element
of $\wc'$) then 
$h=e^{2a_{\xi}}h_0$ (resp.
$h'=e^{2a'_{\xi'}}h_0$) with $a_{\xi}$ (resp.
$a'_{\xi'}$) being a value (0-jet) in $\Rset$ at $\xi\in\widehat{\mathcal{C}}$ of \textit{germs\/}  of $C^2$ fonctions defined on
$\widehat{\mathcal{C}}$, and
$h_0$ being a  metric   at $0$ ($=\varphi_0(p_0)$). Also we easily deduce the following relation between the 0-jets of  metrics $h$ and
$h'$ (which are not metrics fields; See appendix), respectively at $\xi$ and $\xi'$~:
\begin{equation}
h'(\,.\,,\,.\,)=
e^{2(a'_{\xi'}-a_\xi)}h(\,.\,,\,.\,)\,.
\label{primast}
\end{equation} 
Moreover we assumme
that the values for the $a$'s are bounded on $\widehat{U}_0$ (which is relatively compact since $U(p_0)$ is relatively compact and $\varphi_0$ is a proper
map) and such that to $p\in U(p_0)$ corresponds $a_\xi\equiv a_p$ with $\xi=\varphi_0(p)$ (the
same with prime marks). Also, in full generality, at $0=\varphi_0(p_0)$, 
$a'_0\neq a_0$. Then $a_\xi\equiv a_p$ and $a'_{\xi'}\equiv a'_{p'}$ are 0-jets of germs of $C^2$ functions defined on $C$ and $C'$. Also, let us remark
that $\varphi_0$ is not necessary a conformal map.\par We denote by a
dot ``$\,\,\,\dot{\,}\,\,\,$" the derivatives with respect to $s$ or
$s'$.
Then we
assume that (the two worldlines $\wc$ and $\wc'$ are supposed to be in the futur
cones, i.e.  $h>0$ and $h'>0$ when on the two worldlines tangent spaces)
\begin{equation}
h_0(\dot{\xi},\dot{\xi})=1\,,\qquad h_0(\dot{\xi}',\dot{\xi}')=1\,,
\label{normo}
\end{equation}
regardless of the $s$  and $s'$ values, where we denote 
\begin{equation}
\dot{\xi}\equiv\frac{d\wg}{ds}
\equiv\frac{d\xi}{ds}\equiv T_s\wg\,,
\qquad\dot{\xi'}\equiv
\frac{d\wg'}{ds'}\equiv\frac{d\xi'}{ds'}\equiv T_{s'}\wg'\,.
\label{vecdot}
\end{equation}
Then the conditions (\ref{normo}) necessarily involves that there
are no critical points justifying the previous assumption in the definitions of $\gamma$ and $\gamma'$  as embeddings. Moreover we just need there
are of class $C^2$ and, as a consequence, the 
angular discontinuities are cancelled out from the not-piecewise path embeddings assumption.\par
We define the 4-velocity tangent vectors
$\xi^1\in T_\xi\wc$ and
$\xi'^1\in T_{\xi'}\wc'$ by the
relations
\[h(\xi^1,\xi^1)=1\,,\qquad\mbox{and}\qquad h'(\xi'^1,\xi'^1)=1\,,\] 
denoting:
\[
\xi^1\equiv\frac{d\xi}{d\tau}\,,\qquad\xi'^1\equiv
\frac{d\xi'}{d\tau'}\,,
\]
where $\tau\in[0,T]$ and $\tau'\in[0,T']$ are the proper times on $\wc$ and $\wc'$.
Then, since $\tau$ and $\tau'$ are new curvilinear coordinates, then the velocity vectors are colinear and we
find easily that:
\begin{equation}
\dot{\xi}=e^{a_\xi}\,\xi^1\,,\qquad
\dot{\xi'}=e^{a'_{\xi'}}\,\xi'^1\,,
\label{xiu}
\end{equation}
and consequently
\begin{equation}
d\tau=e^{a_{\xi}}\,ds\,,\qquad d\tau'=e^{a'_{\xi'}}\,ds'\,.
\label{times}
\end{equation}
Let us remark that $\xi^1$ and $\xi'^1$ are vector values of  \textit{germs\/}, i.e, 0-jets at $\xi$ and $\xi'$, of velocity 4-vectors fields since the $a$'s
are values of germs of scalar fields themselves. Also we denote by $\|\,\,\|$ (resp.
$\|\,\,\|'$ and ${\|\,\,\|}_0$)
the norm associated to $h$ (resp. $h'$ and $h_0$). \par\medskip   Since $\tot$ is simply connected, it
always exists a local
homeomorphism\footnote{The 1935 Whitney's Theorems assert that any proper continuous map from a
$k$-dimensional manifold to a $n=2k+1$ one, is homotopic to an embedding, and two embeddings from a $k$-dimensional to a $n=2k+2$ one, are isotopic. Hence,
the case $n=4$ is the limit case for our discussion about isotopic embedded paths in manifolds.} (assumed to be of class $C^2$) on $\wc$, denoted
by $f$ such as $f(\wc)=\wc'$, $f(0)=0$ and $f(\iota)=\iota$. This
application $f$ is a homotopy map assumed to keep paths orientations. More exactly it exists a
proper differential map (an homotopy)  $F:[0,1]\times\wc\longrightarrow \tot$ such as
$F(0,\wc)=\wc$ and $F(1,\wc)=\wc'\equiv f(\wc)$. Hence, there exists
a local diffeomorphism between the two projected (on
$\tot$)  worldlines $\wc$ and $\wc'$, as well as between the two
corresponding primary ones on $\m$.  Then, we easily deduce
that $\forall s\in [0,1]$,
$\exists s'\in [0,1]$ such as $f\circ\wg(s)=\wg'(s')$.  \par
Then
from $T_{\wg}f\circ T_s\wg\,ds=T_{s'}\wg'\,ds'$
$\Longleftrightarrow$ $T_{\xi}f\circ\dot{\xi}\,ds=\dot{\xi'}\,ds'$
and (\ref{normo}), we deduce 
\begin{equation}
ds'=|\det(T_{\xi}f)|^{1/4}\,ds\,.
\label{dss}
\end{equation}
Then, if  $|\det(T_{\xi}f)|=1$, meaning $f$ is an isometry, i.e.
an element of the so-called Poincar{\'e} pseudogroup and therefore 
up to an additive constant we can set
$s=s'$.
Moreover we deduce in that case:
\begin{equation}
\dot{\xi}'=T_\xi f\,.\,\dot{\xi}\,,\qquad
\xi'^1=e^{(a_\xi-a'_{\xi'})}T_\xi
f\,.\,\xi^1\equiv\Omega_{\xi',\xi}\,.\,\xi^1\,.
\label{gropi}
\end{equation}
Then $(\xi,\xi'=f(\xi),\Omega_{\xi',\xi})$ is an element of a
conformal groupoid associated to $h$ (or $h'$):
\[
h'(\Omega_{\xi',\xi}\,.\,\xi^1,\Omega_{\xi',\xi}\,.\,\xi^1)
=h(\xi^1,\xi^1)\Longrightarrow
h(\Omega_{\xi',\xi}\,.\,\xi^1,\Omega_{\xi',\xi}\,.\,\xi^1)=
e^{2(a_{\xi}-a'_{\xi'})}h(\xi^1,\xi^1)=e^{2(a_{\xi}-a'_{\xi'})}\,,
\]
from which we also deduce that $T_\xi f$ is a Lorentz transformation for the metric $h$.\par
In fact, in order to modify the
determinant of $f$, considering the couple $(\mathcal{C},\mathcal{C}')$ being fixed and given, we
have in fact a class of admissible couples $(\varphi_0,f)$ which are
equivalent, i.e. meaning that $(\varphi_0',f')\simeq(\varphi_0,f)$ if
and only if $\varphi_0^{-1}\circ f\circ\varphi_0(\mathcal{C})=\mathcal{C}'$
and ${\varphi'}_0^{-1}\circ f'\circ\varphi_0'(\mathcal{C})=\mathcal{C}'$, or
equivalently $\varphi_0^{-1}\circ f\circ\varphi_0(\mathcal{C})={\varphi'}_0^{-1}\circ
f'\circ\varphi_0'(\mathcal{C})$, taking care that the latter is not an equality between functions but image sets. Then, $f$
can be modified  to be an isometry, when $\varphi_0$ is ``adequately"
deformed to $\varphi'_0$.\par
In fact, we can set the relations~: $\varphi_0(\mathcal{C})=\widehat{\mathcal{C}}$ and
$\varphi'_0(\mathcal{C})=\widehat{\mathcal{C}}$, and we also define $f'(\widehat{\mathcal{C}})\equiv\widehat{\mathcal{C}}''$ (see Figure 4.).
Consequently $\varphi'_0\circ\varphi_0^{-1}(\widehat{\mathcal{C}}')=\widehat{\mathcal{C}}''$. Thenceforth, from the relation \eqref{dss},
we deduce after integration that there exists a function $\alpha_f$ such that $s'=\alpha_f(s)$, with $\alpha_f(1)=1$ and $\alpha_f(0)=0$.
Then, we  can redefine the curvilinear coordinate on  $\widehat{\mathcal{C}}'$ with $s$
such that $\widehat\gamma'(s')=\widehat\gamma'\circ\alpha_f(s)\equiv\widetilde\gamma'_f(s)=\xi'$.
From the latter we deduce that
\[
h_0\left(\frac{d\xi'}{ds},\frac{d\xi'}{ds}\right)=|\det(T_{\xi}f)|^{1/2}\neq 1\,.
\]
On the other hand
\[
\frac{d\xi''}{ds}=T(\varphi'_0\circ\varphi_0^{-1})\,.\,\frac{d\xi'}{ds}\,.
\]
And, if we consider that
\[
h_0\left(\frac{d\xi''}{ds},\frac{d\xi''}{ds}\right)=1\,,
\] 
finally one deduces the relation which allows us to obtain $\varphi'_0$ from
$\varphi_0$ in a conformal way~:
\[
|\det\,T(\varphi'_0\circ\varphi_0^{-1})|=|\det\,T_{\xi}f|^{-1}\,,
\]
both with $\widehat{\mathcal{C}}''$ being parametrized by $s$ instead of $s''$ (see Figure 4.), 
and which involves that the absolute value of the determinant of
$Tf'$ equals 1. Then, we consider hereafter $f$ to be an isometry.
\par\medskip
Then, there are two main classes of
physical assumptions, closely related to the elevator
metaphor: the case for which physical
considerations are made somehow in reference to the
paths intersection event at $p_0$, and the other
considering the viewpoints of  travelling observers on
$\mathcal{C}$ or $\mathcal{C}'$. Also, we will show that
the terms $a_\xi$ are, up to a constant for units,
potential of accelerations and thus, that their variations are
related to accelerations vectors. It will follow,
as shown below, the first case corresponds to what we
call the
\textit{``all-motion/all-gravitation"\/} assumption,
whereas the second corresponds to the 
\textit{``intermediate"\/} one.
\par\medskip
\underline{\textbf{Preliminaries:}} 
We denote by $\eta$ any vector in $\tot$, and then  from the 0-jet feature of $a_\xi$, we set below the definition 
of $b_\xi$ as the \textit{``gradient-like"\/} vector (evaluated from \textit{germs\/} of scalar fields with value $a_\xi$ at $\xi$),  
with respect to the 0-jet of metric $h$ at $\xi$~:
\begin{equation}
da_\xi\equiv da_\eta{\big/_{\!\!\xi}}\equiv h(b_\xi,d\eta{\big/_{\!\!\xi}})\,,
\label{dab}
\end{equation}
(taking care that between values of germs at $\xi$, we get the general situation~: $b_\xi\not\simeq\vec \nabla a$, since $a$ is a 0-jet at $\xi$ and not a
function; Nevertheless we can have the equality along the worldline since $a_\xi$ can be considered as a function $a_\xi\equiv\alpha(\tau)$ and
$b_\xi\equiv\beta(\tau)$ the given gradient of $a_\xi$ along this same worldline) and we denote
\[
b^0_\xi\equiv e^{2a_\xi}b_\xi\,,
\]
(the same with prime marks). From (\ref{xiu}) and denoting
$\xi^2=d\xi^1/d\tau$, we  obtain also in addition the relation:
\begin{equation}
\xi^2=e^{-2a_\xi}
\left\{
\ddot{\xi}-h_0(b^0_\xi,\dot{\xi})\,\dot{\xi}
\right\}\,.
\label{xid}
\end{equation}
We have to take care of that $h_0(\ddot{\xi},\dot{\xi})=0$ whereas
$h(\xi^2,\xi^1)\neq0$ in general.
\par\medskip
We denote by 
$\nabla$  (resp. $\nabla^0$)  the Levi-Civita covariant derivative at any $\eta\in\widehat{U}_0$ (resp. at 0) associated to any given representative metric
field of germs   with the 0-jet metric $h$ at $\xi$ (resp. $h_0$ at $0$). 
Then, we deduce from the definition of $\nabla$, and when restricted on the paths and not at angular discontinuities, $\forall\,\,\rho,\,\zeta$ being vector
values (0-jets) at $\xi$ of germs of vector fields on $\widehat{\mathcal{C}}$~:
\begin{equation}
{\nabla}_{\rho}\zeta={\nabla}^0_{\rho}\zeta+
h(b_\xi,\rho)\zeta-h(\zeta,\rho)b_\xi+
h(b_\xi,\zeta)\rho\,.
\label{covdev}
\end{equation}
And also from the definition of the Levi-Civita covariant derivative
\begin{equation}
h({\nabla}_{\zeta}\zeta,\zeta)=\zeta.h(\zeta,\zeta)\,.
\label{zeta3}
\end{equation}
In particular, we deduce all along $\widehat{\mathcal{C}}$~:
\begin{equation}
{\nabla}_{\xi^1}\xi^1=e^{-2a_\xi}
\left\{
{\nabla}^0_{\dot{\xi}}\dot{\xi}+
h_0(b_\xi^0,\dot{\xi})\dot{\xi}-b_\xi^0
\right\}\,.
\label{haant}
\end{equation}
This is a Haantjes vector (see the line ``b)" in formula (15) in
\cite{haantjesconf41}), which is conformally equivariant, i.e. $\nabla_{\xi^1}\xi^1$ is transformed to $\nabla'_{\xi'^1}\xi'^1$ by $\Omega_{\xi',\xi}$.
Let us remark the important fact that if we set the so-called \textit{meshing assumption\/} of Ghins and Budden \cite{ghinsbuden2001}~:
\begin{equation}
{\nabla}^0_{\dot{\xi}}\dot{\xi}\equiv\ddot{\xi}\,,
\label{derab}
\end{equation}
a colinearity which is possible from (\ref{normo}) and 
$h^0({\nabla^0}_{\dot{\xi}}\dot{\xi},\dot{\xi})=\dot{\xi}.h^0(\dot{\xi},\dot{\xi})=\dot{\xi}(1)=0$. Then from
(\ref{xid}) we find on $\widehat{\mathcal{C}}$:
\begin{equation}
{\nabla}_{\xi^1}{\xi^1}=
\xi^2 +2h(b_\xi,\xi^1)\xi^1-b_\xi\,.
\label{covac}
\end{equation}
We will see that ${\nabla}_{\xi^1}{\xi^1}$ can be
attributed to an acceleration given in the proper frame by mechanic
gauges like our human bodies when ``feeling" accelerations\dots whereas the
$\xi^2$ alone is a measured acceleration of motion with respect to another
external point in the vicinity of $\xi$, and thus having a \textit{conventional\/}
status as it will be precised in the sequel in recalling that the covariant 
derivatives are also conformally equivariant vectors on the contrary to the
$\xi^2$'s. We can also notice that:
\begin{equation}
h(\nabla_{\xi^1}\xi^1,\xi^1)=0\,,
\label{hnab}
\end{equation}
from which we deduce with the relation \eqref{covac} that
\begin{equation}
h(\xi^2,\xi^1)=-h(b_\xi,\xi^1)\qquad\mbox{($\neq0$ in full generality)}\,.
\label{hx2hb}
\end{equation}
We call $\nabla_{\xi^1}\xi^1$ the \textit{``gauged acceleration"\/}, i.e. the acceleration gauged by its velocity $\xi^1$.
\par\medskip
\underline{\textbf{Case 1 :}}
(\textit{``all-motion/all-gravitation"\/} assumption) 
Then, we have the formula ($\|\xi^1\|>0\Longrightarrow$ the signature of $h_0$, $h$ and $h'$ is $(+,-,-,-)$\,):
\begin{equation}
\ln\|\xi^1\|=
a_{\xi}+\ln{\|\xi^1\|}_0=0\,,
\label{scale}
\end{equation}
meaning that at $\xi$ any potential of acceleration with value $a_\xi$ assumed to be due to
gravitation,  can also be considered as, or equivalently as, having a $\xi^1$ velocity
origin. It involves also from (\ref{scale}) that scaling might be due to
velocity, i.e. a lenght contraction phenomenon in special relativity.
\par\medskip  
Also from the Haantjes vector (\ref{covac}), if
${\nabla}_{\xi^1}\xi^1$ is set to
$0$ (we consider a geodesic, i.e. special relativity), then we obtain:
\begin{equation}
\xi^2=
b_\xi-2h(b_\xi,\xi^1)\,\xi^1\,.
\end{equation}
And, if the motion, i.e. $\xi^1$, is normal to the acceleration vector of
gravity $b_\xi$, then $\xi^2\equiv b_\xi$ and $h(\xi^2,\xi^1)=0$. In other words,
there is no gravitation but only motion, or no motion and
gravitation only; All is motion or exclusively all is gravitation. But
also it shows that vanishing of ${\nabla}_{\xi^1}\xi^1$
is the indicator for free-falling  (geodesic) motions, inducing a local special relativity, 
as demonstrated again by Ghins and Budden
\cite{ghinsbuden2001}.
\par\medskip
Also we add (and it is important for the sake of latter physical interpretations) that $\xi'^1$ (or $\xi^1$ anyway) has the status of a
``transported relative velocity" between two travellers at $\xi'$
and $\xi$. Indeed, at first, it is a relative
velocity as it suffices to verify at
$s=s'=0$. For that, denoting by $\eta'^1$ (resp.
$\eta^1$) the vector
$\xi'^1$ (resp. $\xi^1$) at $0$ ($=\varphi_0(p_0)$) with
${\|\eta'^1\|}_0={\|\eta^1\|}_0=1$, and $T_{\xi}f$
being the tangent map of $f$ at $\xi$, then we easily deduce if $a'_0=a_0$
that $T_0f$ is a  matrice of isometry (i.e. a Lorentzian one in the pseudo-Euclidean case) and that
$\eta'^1=T_0f.\eta^1$ is a relative velocity vector
since $\eta^1$ would be the time-like tangent vector
attached to $\gamma$ at $p_0$. And, secondly, this
vector $\eta'^1$ is transported along
$\wg'$ when considering $s'\neq0$. Hence, the $f$ isometric homotopy involves
that the Lorentz group is an holonomy group in the present
case. Then, we consider hereafter and throughout the paper that $a'_0=a_0$.\par\medskip
\underline{\textbf{Case 2 :}}
(\textit{``intermediate"\/} assumption) Coming back to the formula
(\ref{scale}), we ask for how an observer, in his own proper
frame, can evaluate a vector such as $\xi^1$ ? Indeed, in writing
``$\xi^1$", we mean a 4-velocity at $\xi$ but with respect to which observable
point in the closed vicinity of $\xi$ ?! This would be the case with respect to the initial
crossing point $0$ if still ``observable" in the future meanwhile by definition this event doesn't exist any more.  In fact from an
experimental point of view, since an absolute reference point
doesn't exist, $\xi^1$ ``appears" to be arbitrary while it is not \cite[see appendix II]
{bergsondurP,bergsondurS,bergsondurEn}~! The frames ``appear" to become relative themselves as well as
the vectors (!),  which is the ``essence" of the relativity
theory.\par  As a consequence, the observer can split the gravitational part and the velocity one
in such a way that the relation (\ref{scale}) is always maintained.
That means  he could consider
$\tilde a_{\tilde{\xi}}\equiv a_\xi+\upsilon_\xi$, possibly at an other point $\tilde{\xi}$ with velocity $\tilde{\xi}^1$, and
$\ln{\|\tilde{\xi}^1\|}_0\equiv\ln{\|\xi^1\|}_0-\upsilon_\xi$  as
other admissible potentials and velocities with an other potential of
gravity $\upsilon_\xi\equiv\upsilon_\xi(\tilde{\xi})$ being \textit{a priori\/} an arbitrary value. Hence what would be really
assessed as an \textit{intermediate situation\/}, could be only the relative
discrepancy, i.e. $\upsilon_\xi\equiv \tilde a_{\tilde{\xi}}-a_\xi$, or 
$\upsilon_\xi\equiv \ln({\|{\xi}^1\|}_0/{\|\tilde{\xi}^1\|}_0)$, or (up to an arbitrary multiplicative constant)~: 
\begin{equation}
\frac{1}{3}\,\upsilon_\xi\equiv (\tilde
a_{\tilde{\xi}}-\ln{\|\tilde{\xi}^1\|}_0)-(a_\xi-
\ln{\|\xi^1\|}_0)=\tilde
a_{\tilde{\xi}}-a_\xi-\ln({\|\tilde{\xi}^1\|}_0/{\|\xi^1\|}_0)\,,
\label{upsilon}
\end{equation}
the definition depending on the experimental protocol for evaluation and/or the convention of splitting we  adopt in the latter
protocol to discriminate between velocity and potential of gravity origins (thanks to the Einstein elevator metaphor translated in the potentials of
acceleration case rather than accelerations). Let us remark that the last  definition for
$\upsilon_\xi$ is  analogous to (\ref{covac}) with the difference between $\xi^2$ and
$b_\xi$. Roughly speaking, the normalization of $\xi^1$ expressed by the relation \eqref{scale}, won't be experimentaly obtained, but the relative scalar
$\upsilon_\xi$ only.\par It matters that if the measurements of the variations of $\upsilon_\xi$ reflect those of $\tilde
a_{\tilde{\xi}}$ and $\ln{\|\tilde{\xi}^1\|}_0$ satisfying an analogous relation \eqref{scale}, then a increasing of $\tilde{a}_{\tilde{\xi}}$ to 0
(considering a negative attractive potential at
$\xi$) when $\tilde{\xi}-\xi$ tends toward infinity, involves a decreasing of ${\|\tilde{\xi}^1\|}_0$ and thus, an increasing of the modulus of its 3-vector
part, thanks to the signatures of the metrics. The latter modulus increasing could make objetcs at $\xi$ and $\tilde\xi$ look further away. This could be 
related to the leak of galaxies and the Hubble constant would have to be taken into account in the definitions of the potentials $a$. On the other hand,
this modulus variation could be related to object moves bringing them closer together. To discriminate between these two options, we can evoke the Unruh paper
\cite{unruham81} and his discussion on the twins paradox based on principles of stereometric measurement in experimental protocols. As he clearly
demonstrated, a change of acceleration is \textit{relative\/}, i.e. if a twin takes the decision to accelerate and stop to be in an inertial frame, or is
suddenly embedded in a gravitational field (thanks to the Einstein elevator metaphor !), then his vectorial acceleration ``feel" is exactly the vectorial
opposite of the kinematical acceleration of the other twin, as in a kind of action-reaction principle. In other words, and in our formalism, if $b_\xi$
becomes suddenly non-vanishing whereas
$\nabla_{\xi^1}\xi^1=0$ and $h(b_\xi,\xi^1)=0$ all along this variation (geodesic motion and/or special relativity), then $b_\xi=\xi^2=-\tilde{\xi}^2$, i.e.
each  twin ``sees" the other to be suddenly moving away if $b_\xi$ is directed in the $\tilde\xi$ to $\xi$ direction (let us recall that the vectors $\xi^i$
are evaluated with respect to a point, which could be $\tilde\xi$, but different from $\xi$). Hence, a variation 
$\triangle b_\xi$ is the opposite of a variation
$\triangle\tilde{\xi}^2$. Then, passing from $\xi$ to $\tilde{\xi}$ with $b_\xi$ decreasing and directed all along in the $\tilde{\xi}$ to $\xi$ direction,
means $\triangle b_\xi$ at $\tilde\xi$ is directed in the
$\xi$ to $\tilde{\xi}$ direction, and then $\triangle\tilde{\xi}^2$ at $\xi$ is directed in the $\tilde{\xi}$ to $\xi$ direction, meaning we have a leak. 
\par\medskip  But in fact, without any
other point of reference such as
$\tilde{\xi}$, the velocity
$\xi^1$ could be only deduced by integration from the vector
${\nabla}_{\xi^1}{\xi^1}$ which can always be obtained from local mechanic gauges without need of distance measurements. But a constant of  integration
would remain and since it would be deduced by integration,  from recorded measurements of the mechanic gauges states, it
would refer to a kind of ``mean" frame, and not to the moving frame at which the recordings finish, which is the one containing the right velocity
vector $\xi^1$. Hence the result would be purely formal or virtual though less arbitrary or relative, leading to 
\textit{``virtual velocities $\V^1$"\/}. This has to be linked strongly to the concept of
\textit{``phantasmal or virtual viewpoint"\/} defined by Bergson \cite[see appendix
III]{bergsondurP,bergsondurS,bergsondurEn}, and also to the convention involved in defining spacetime coordinates in the special relativity framework as W.
Unruh demonstrated again \cite{unruham81}.\par\medskip Also from
\eqref{gropi} and (\ref{covac}), we deduce
\begin{equation}
\xi'^2
+2h'(b_{\xi'},\xi'^1)\xi'^1-b_{\xi'}=\Omega_{\xi',\xi}\left\{\xi^2
+2h(b_\xi,\xi^1)\xi^1-b_\xi\right\}\,.
\label{claw}
\end{equation}
Also, clearly, this shows the affine feature of the
acceleration of gravitation  vectors when passing
from a path to another path. This is again close to
the Einsteinian relativity and to the Einstein's
principle of equivalence. Indeed the elevator
metaphor in this principle, points out the affine
feature of the acceleration which involves
equivalence between relative uniformly accelerated
frames when the latter have a relative velocity
$\xi'^1$ ($\tau'=0$) and relative acceleration
$\xi'^2$ ($\tau'=0$) at their crossing point $p_0$.
\subsection{Consequences on physical interpretations}
\subsubsection{Integrations along worldlines and a conformal Langevin
twins paradox solution}
In full generality, the time discrepancy at $\iota$ is defined by
the relation (where $\wl\equiv\wc'\circ{\wc}^{-1}$)~:
\begin{equation}
\triangle_{\iota}(\wl\,)=
\int_0^1\!h'(\xi'^1(s),\xi^1(s))\,d\tau'(s)-
\int_0^1\!h(\xi'^1(s),\xi^1(s))\,d\tau(s)=
\int_0^1\!h(\xi'^1(s),\xi^1(s))\,\left(e^{3(a'_{\xi'(s)}-a_{\xi(s)})}-1\right)d\tau(s)\,,
\label{timetwintau}
\end{equation}
where $a'_{\xi'(s)}$ and $a_{\xi(s)}$ are scalar sections from $[0,1]\ni s$ to $\Rset$ which are not \textit{``factorizable"\/}, meaning it doesn't  exist 
scalar fields $A'$ and $A$ from $\tot$ to $\Rset$ such as $a'_{\xi'(s)}\equiv A'\circ\hat{\gamma}'(s)$ and $a_{\xi(s)}\equiv A\circ\hat{\gamma}(s)$ (the
same for the velocity vectors). This remark forbids, by pull-back by $\hat\gamma$ (and $\hat\gamma'$), to consider equalities such as
\[
\int_0^1\!h(\xi'^1(s),\xi^1(s))\,d\tau(s)\equiv\int_{\widehat{C}'\times\widehat{C}}\,h(\xi'^1(\xi'),\xi^1(\xi))\,{}^\ast d\tau(\xi)\,,
\] 
and then pulled back 1-forms (fields of co-vectors) $h(\xi'^1(\xi'),\xi^1(\xi))\,{}^\ast d\tau(\xi)$ on $\tot\times \tot\ni(\xi,\xi')$.

Or equivalently, it forbids to write (${\wl}^{-1}\equiv\wc\circ{\wc}'^{-1}$)~:
\begin{equation}
\triangle_{\iota}(\wl\,)
\equiv\oint_{\wl\times{\wl}^{-1}}\omega\,,
\label{timetwincercle}
\end{equation}
where $\omega$ is a  1-forms  along the loop equals to  $h(\xi'^1(\xi'),\xi^1(\xi))\,{}^\ast d\tau(\xi)$ on
$\widehat{C}\times\widehat{C}'$ and
$h'(\xi'^1(\xi'),\xi^1(\xi))\,{}^\ast d\tau'(\xi')$ on $\widehat{C}'\times\widehat{C}$.\par
A few comments are needed to explain our definition of $\triangle_{\iota}(\wl\,)$, and precisely the definitions of the two integrals in \eqref{timetwintau}.
Let us define $\delta\tilde{\tau}\equiv\gamma_0\,e^{a_\xi}\,ds$ where $\gamma_0\equiv h_0(\xi'^1,\xi^1)$. We can consider
$\gamma_0$ as the ``gamma" coefficient usually used in special relativity, and then, at a first approximation, we write
$\gamma_0\simeq1-\frac{1}{2}\left(\frac{\tilde{u}}{c}\right)^2$ where, at first approximation again, $\tilde{u}$ is the modulus of the relative 3-vector of
velocity associated to $\xi'^1$, and $c$ the speed of light. Then we deduce the following approximation~:
\[
\delta\tilde\tau-ds\simeq \left(a_\xi-\frac{\tilde{u}^2}{2c^2}\right)\,ds\,.
\]
It follows that considering the following ascriptions~: $\delta\tilde\tau\leftrightarrow \tau$ (``the  time recorded by the flight"),
$ds\leftrightarrow \tau_0$ (``the ground reference time") and at a first approximation~: $a_\xi\leftrightarrow$ \textit{the potential of gravity\/}
\cite{hafelekeat172}, we obtain exatly the formula given by J.~C. Hafele and R. Keating when they computed the theoretical infinitesimal time discrepancy
between two flying atomic clocks
\cite[formula (2)]{hafelekeat172}:
\[
\tau-\tau_0=[gh/c^2-(2R\Omega\, v+v^2)/2c^2]\,\tau_0\,,
\]
where $R$ is the earth radius, $\Omega$ the angular velocity of the earth and $v$ the velocity of the flight with respect to the ground frame on earth, $h$
the altitude of the flight and $g$ the acceleration of gravity constant.\par 
In fact the Hafele and Keating experiment and the latter ascription would require to assume and to consider a third trajectory denoted
$\widehat{C}_0$ from 0 to $\iota$ for the ``fixed" (on earth) ground frame with the metric $h_0$ all along, i.e. a third embedding
$\hat{\gamma}_G\equiv\xi_G$, with $\tilde u$ being the relative velocity with respect to this ground frame (see Figure 5. below). 
\begin{center}
\noindent\fbox{\includegraphics[scale=.6]{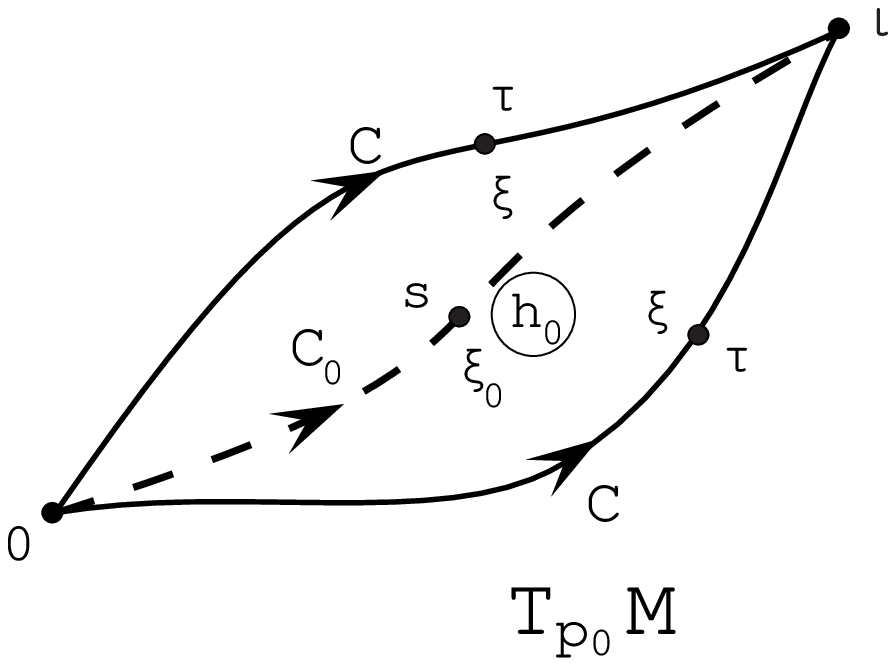}
Figure 5.}
\end{center}
Then we would consider two time discrepancies~:
one between $\widehat{C}$ and $\widehat{C}_0$, and the other between $\widehat{C}'$ and $\widehat{C}_0$. Then we would substract them in such way that we
would recover our times discrepancy definition for which just two paths have to be taken into account, namely $\widehat{C}$ and $\widehat{C}'$.\par Another
argument would be that such formula expresses a ``kind" of transport (from 0 to
$\xi$ and $\xi'$) of the parallax effects of the special relativity such as the time dilatation.\par Also, a very fundamental consequence of
\eqref{timetwintau} is that proper times in general relativity do not have the same definition as in special relativity. They must be also ``relative".
Indeed, if the $\tau$'s are proper times coming up and defined from special relativity, the $\tilde\tau$'s are those times which would have to be considered
and defined as the relative proper times of the general \textit{and\/} special relativity; ``Relative" because of the terms $\gamma$ and $\gamma'$ depending
on three vectors ($\xi^1$, $\xi'^1$, and $\xi'$ or $\xi$) whereas the $\tau$'s depend on one vector only ($\xi$ or $\xi'$). With such definition, we insure a
continuity between general and special relativity. But also, and as very fundamental point, general relativity must deals with metrics such as
$d\tilde\tau^2$, defined in the case of factorizable embeddings on
$T\tot\times T\tot$ or $\tot\times\tot$ in the case of given sections; And consequently, spacetime would be defined by metrics on
$\tot\times\tot\simeq\m\times\m$ which would the \textit{Spacetime\/} of the general relativity and not $\m$.\par As a result of our definition for the time
discrepancy, 
$\triangle_{\iota}(\wl\,)$ is not necessarily vanishing, unless the relative variation of 0-jets of potential $\triangle{a}\equiv{a'_{\xi'}-a_{\xi}}$ equals
zero all along the paths. Let us remark that the difference of the two integrals can 
be merged into one only because $f$ is an isometry, i.e. a Lorentz transformation. It matters to notice that, although $U(p_0)$
is simply connected, the loop integral which could define
$\triangle_{\iota}(\wl\,)$ will not be necessarily vanishing (!) even
if in all cases the loop $\wl$ is contractible (homotopic to the
point $0$). In fact, we can't do this contraction in the loop integral, since it would require the knowledge and the unicity of a particular field of a
continuous 1-form  to allow the loop contraction in the whole of
$\widehat{U}_0$ whereas we only have co-vector values   $\delta\tilde\tau$ of germs of 1-forms,  i.e. a
\textit{local ring\/} of 1-forms. Nevertheless,  if we have possibly 1-forms $\delta\tilde\tau$, the latter would be ``dilatated" by the pull-back defined from
the loop contraction and it could not necessarily reduce the time descripancy to 0. Then a variation of
$\omega$ on the loop involves a discrepancy of time ``betwins". Thus, it rules out the necessity of introducing a peculiar given global topology (homotopy) of
$\m$ since the problem is solved locally on
$U(p_0)$.
\par\medskip
The case for which $\triangle a\equiv0$ constantly along the path (which doesn't need that the $a$'s are potential scalar fields depending on the $\xi$'s,
i.e. factorizable), means there is no relative potential of acceleration (there may have non-vanishing difference of potentials of gravity since
the $a$'s are not potentials of gravity but of acceleration only); And then, it involves necessarily that
$\triangle_{\iota}(\wl\,)=0$. That was the result given by H. Bergson in appendix III of
\textit{``Dur{\'e}e et simultan{\'e}it{\'e}"\/} in 1922. But the definition of
$\triangle_{\iota}(\wl\,)$  involves also that a relative potential of acceleration ($\triangle a$) can produce a time discrepancy as A. Einstein noticed
it about the twins paradox~! Roughly speaking, we can consider the potentials of acceleration as kind of ``catalysts" for the special relativistic
contribution to the time discrepancy $\triangle_{\iota}(\wl\,)$.\par\medskip But, as mentioned previously in the case 2 (i.e. the \textit{intermediate\/}
assumption case with
$\tilde{a}_{\tilde\xi}\equiv a'_{\xi'}$), this integral can't be computed by any observer because of generally unknowable 0-jets of potentials $a_\xi$ and
$a'_{\xi'}$.  We must take care in the Hafele and Keating experiment that $a_\xi$ is the potential of gravity at a first approximation, since we must
consider that there is a neglected dependency with respect to the norm $\|\xi'^1\|_0\simeq 1$ coming from the relation \eqref{scale}. Then, we can write  from
\eqref{scale} (again we consider the
$\xi^{(1)}$'s as non-factorizable proper embeddings from
$[0,1]$ to the tangent space of $\tot$, namely $T\tot$; The same holds for the metrics $h$ and $h'$ because of the non-factorizable sections ``$a$")~:
\begin{equation}
\triangle_{\iota}(\wl\,)=\int_0^T\!\!
h(\xi'^1,\xi^1)\,(
\|\xi'^1\|^3\,e^{\upsilon_\xi(\xi')}-1)\,d\tau\equiv T'-T\,,
\label{merged}
\end{equation}
with the $\xi$'s and the $a_\xi$'s satisfying the relation of constraints \eqref{scale}, meaning that we consider non-factorizable proper embeddings from
$[0,1]$ to a submanifold of
$T\tot\times\Rset\ni(\xi,\xi^1,a_\xi)$. We obtain another formula but for which $\xi'^1$ has to be
estimated with respect to $\xi^1$ and $\xi$. Thus, an experimental process must be given to make such
evaluation. That is the subject of the next subsection. \par \medskip
On that point, we give some remark about the computations made in the Hafele and Keating experiment. Indeed, to compute the time discrepancy between the two
aircrafts of the experiment, two separate computations has to be made: one in considering special relativity and the other general relativity, these
computations were not made on the same footing whereas general relativity would include special one ! In fact, it exhibits the fact that general relativity
can't consider relative velocities as special relativity does. On contrary, our formula includes in the same time, positions (usual general relativity)  and
velocity (special relativity) in a unique formalism. The Hafele and Keating computations shows an incompletness of the general relativity to include special
relative effects, and as a consequence, we would have two separate theories ! This successful experiment shows successfully how the special and general
relativities could be viewed as different theories. It is not the case as our time discrepancy formula exhibits it.
\subsubsection{``Two Worldlines" experimental protocols}
\paragraph{The tools}
The Hafele and Keating experiment is what we call a three worldlines experimental protocol and this protocol allows us to avoid the integrals merging in the
definition of the time discrepancy. In fact, we can consider this protocol as  two ``two worldlines" protocols, each one associated to a couple
``aircraft-ground base". In merging the integrals, we consider only one two worldlines protocol, and we can obtain the time discrepancy formula between one
aircraft and the ground base   by subsituting $\tau$ by $s$, $\tau'$ by $\tau$, $h'$ by $h$ and $h$ by $h_0$ in the formula \eqref{merged}. 
But the great difficulty comes from evaluations of non-approximated values of the fields of gravity and
the velocities. The experimental accord with the theory as been obtained at first approximation, and the Hafele and Keating experiment can't be used, because of
the approximations, in cases of high velocities and fields of gravity. Then, we give, as examples, the following two worldlines protocols to emphasize the
experimental difficulties which can be encountered in obtaining non-approximated experimental values and the resulting undeterminism it could involve.  Of
course, the main difficulty results from the time delay between what each twin observes from the other. It is the first cause of undeterminism  which can't be
cancelled out expect for very specific experimental situations. But the goal is to find protocols giving a deduced non-approximated result in
accordance with the time discrepancy observation at any point on the worldlines and especially at the ending crossing point
$\iota$, which is in fact the meaning of the twins pseudo-paradox resolution. Morevover we consider a peculiar and may be the worth protocol in order to
exhibit at least all kind of experimental difficulties. It is not the simplest one but the one which could reveal the full set of experimental problems.\par 
Thus, we consider an initial time synchronization of some processes at the same spatial position, i.e. at the initial crossing point. And also, constant known
time delay or ahead of time of ``computation" between each algoritmic steps involved in  any embarked travelling or on-earth processes. We would say that we
would have clock processes, but one of the main advantage of the protocols presented below, would be rather to give precise correspondances between spatial
positions and time positions, i.e. a precise evaluation of  the positions events in the spacetime or equivalently a precise deduction of the two
worldlines.\par Hence, these protocols must give for any values of $\tau$ and
$\tau'$, i.e. all along the two worldlines~:
\begin{itemize}
\item The exact values of the spacetime positions $\xi$ and $\xi'$ as well as the velocities $\xi^1$ and $\xi'^1$.
\item The values of $a_\xi$ and $a'_{\xi'}$.
\item Processes of transmission of the latter values between each twin, as well as their corresponding proper times $\tau$ and $\tau'$.
\end{itemize}
We have to focus on the fact that these values don't correspond to the ones observed by the twins when no transmission processes of these latter values exist.
Hence, we consider the \textit{``observed values"\/} we denote by brackets : $[\dots]_{Ob(\xi)}$ or $[\dots]_{Ob(\xi')}$.
\par\bigskip
The  fundamental tools, we will consider, are the covariant derivatives of the velocities~:
\begin{equation}
\nabla_{\xi^1}\xi^1\,,\qquad\nabla'_{\xi'^1}\xi'^1\,.
\label{covdata}
\end{equation}
These are  accelerations ``felt" by the twins, we call the \textit{``travelling" twin\/} and the \textit{``on-earth" twin\/}. These accelerations can be
experimentally evaluated from mechanical gauges in accelerometers, embarked on board with the twins. This is a kind of ``blind" experiment since no knowledge
of the spatial (i.e. geometrical) environment is required by the  twins. Also, these accelerations avoid to define a convention in the splitting of the
accelerations, because of the elevator metaphor, between parts due to  kinematical accelerations and parts due to  accelerations of gravity.\par\smallskip
\vspace{-.5cm}
\paragraph{\normalsize A ``two worldlines" protocol with acceleration measurements.}
In this first protocol, which is the most complex, we consider a process of transmission of the data given by the accelerometers, from the travelling twin to
the on-earth twin. The latter, receiving these data,  geometrically interprets the associated covariant derivative of $\xi'^1$  as
being the same 4-vector but ``relative" to his own proper frame at $\xi$. We call this  4-vector the \textit{``observed covariant derivative at
$\tau$"\/}~:
\[
[\nabla'_{\xi'^1}\xi'^1]_{Ob(\tau)}\,,
\] 
and which differs from the original one as outlined in the figure below (Figure 6.; $\tau'(\tau-c\,\triangle(\tau))$ being the \textit{a priori\/}
unknown value of
$\tau'$ at which the data are transmitted to and received at $\xi(\tau)$)~:
\begin{center}
\noindent\fbox{\includegraphics[scale=.6]{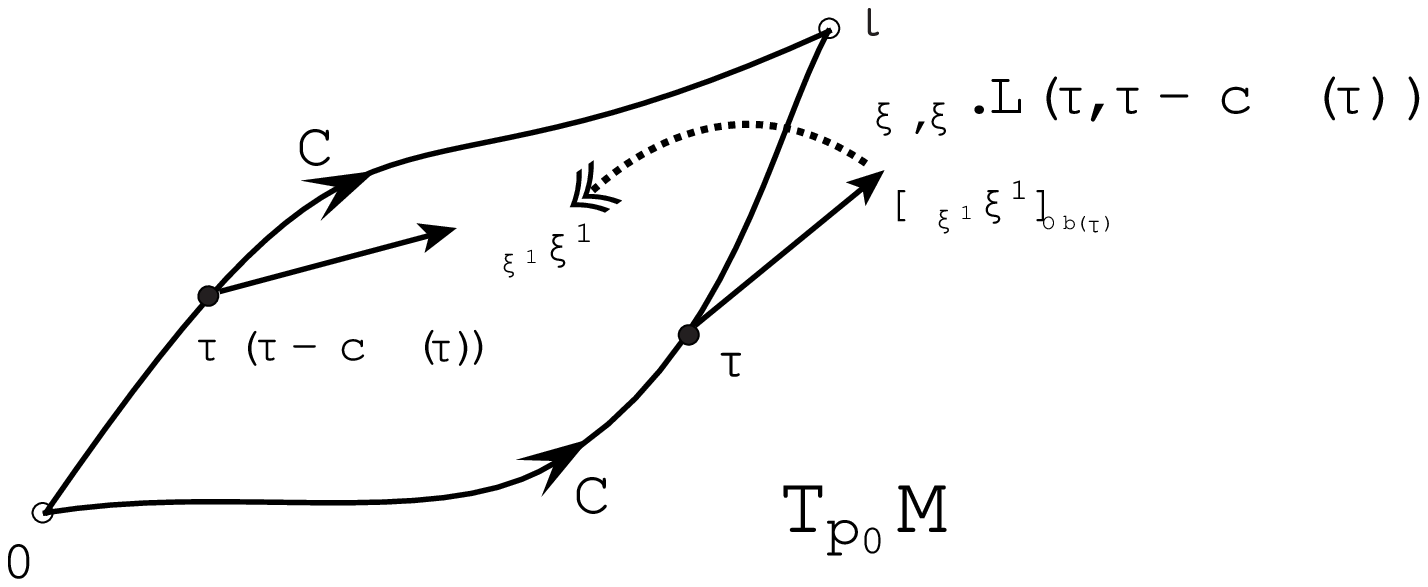}
Figure 6.}
\end{center}
Roughly speaking, the observed covariant derivative is oriented with the same angle with the curve $\widehat{\mathcal{C}}$ than the ``original" covariant
derivative does with the curve $\widehat{\mathcal{C}}'$. Then, from the definition of the time delayed transformation $\Omega_{\xi',\xi}$ at
$\tau-c\,\triangle(\tau)$, with $\triangle(\tau)\equiv[\triangle(\tau)]_{Ob(\xi)}$  being the \textit{observed\/} spatial distance between the twins
evaluated by the on-earth twin at
$\tau$ (or $\xi$), thanks to his  clock, and that we denote by
$\Omega_{\xi',\xi}^\dashv$, we deduce that the covariant derivative at $\xi'(\tau')$ is~:
\begin{equation}
\nabla'_{\xi'^1}\xi'^1=\Omega_{\xi',\xi}^\dashv\,.\,L(\tau,\tau-c\,\triangle(\tau))\,.\,[\nabla'_{\xi'^1}\xi'^1]_{Ob(\tau)}\,,
\end{equation}
where $L(\tau,\tau-c\,\triangle(\tau))$ is the parallel transport from  $\tau-c\,\triangle(\tau)$ to $\tau$.\par
Then, by integration of this acceleration with respect to the proper time of the on-earth twin, we can deduce the velocity $\xi'^1$ at $\xi'(\tau')$
(corresponding to $\xi$ at $\tau-c\,\triangle(\tau)$)~:
\begin{equation}
\xi'^1=\xi'^1_0+\int_0^{\tau}\!\!\Omega_{\xi',\xi}^\dashv\,.\,L(u,u-c\,\triangle(u))\,.\,[\nabla'_{\xi'^1}\xi'^1]_{Ob(u)}\,du\,,
\end{equation}
with $\xi'^1_0$ being the velocity at 0. But, from the definition of $\Omega_{\xi',\xi}^\dashv$, we see that the latter formula is in fact an integral equation
with respect to $\xi'^1$~:
\begin{equation}
\xi'^1=\xi'^1_0+\int_0^{\tau}\left\{\frac{e^{-\upsilon_\xi/3}}{\|\xi'^1\|}\,T_\xi
f\right\}_{\!u-c\,\triangle(u)}\!\!\!\!.\,L(u,u-c\,\triangle(u))\,.\,[\nabla'_{\xi'^1}\xi'^1]_{Ob(u)}\,du\,,
\end{equation}
and which has to be solved to provide a deduced time discrepancy at the ending crossing  point $\iota$, but also for any given value of $\tau$. In this
integral equation, the tangent map $T_\xi f$ is also depending on and defined univoquely by the velocity $\xi'^1$ from \eqref{gropi}, but
$\upsilon_\xi(\xi')$ must be experimentally evaluated or deduced. This latter can be experimentally evaluated by the on-earth twin, and simultaneously
with the covariant derivative data reception,  in computing the potential field of gravity
$\upsilon_\xi(\xi')$ deduced from the known masses distributions or deduced by susbtracting to frequency  shifts of
spectral light rays (for instance) those due to relativistic Doppler effects and then from evaluating $\upsilon_\xi(\xi')$ from the resulting frequency
shifts at rest only due to gravity fields due to the masses distributions.\par As Unruh showed in his paper on the twins paradox, the latter geometrical
evaluations of the
\textit{observed\/} positions $[\xi']_{Ob(\xi)}$ and the
\textit{observed\/} velocities $[\xi'^1]_{Ob(\xi)}$ which are necessary to deduce the relativistic Doppler effects for the evaluation of
$\upsilon_\xi(\xi')$($\neq\upsilon_\xi([\xi']_{Ob(\xi)})$), can be done by telemetry for instance, but other meanings exist, and that it involves spatial
position and velocity measurements with respect to a ``virtual" or ``conventional" system of coordinates, as if the spacetime domain between the on-earth and
the travelling twins would  be a flat spacetime; And then without deviations of the light rays due to a curvature. Hence, the computed
$\xi'^1$ would give a vector with respect to the latter convention of flatness. In fact, it is not a theoretical difficulty as well as an experimental one.
Indeed, it is coherent with the definition of the \textit{observed\/} covariant derivative, since the latter will be interpreted by the on-earth twin precisely as an
acceleration of the travelling twin in a flat spacetime; The latter convention being the simplest since the contrary would need complex or impossible spacetime
curvature measurements all along the transmission of data worldline and moreover simultaneously during the transmission ! In order words, the resulting
computed velocity
$\xi'^1$ will be considered by the on-earth twin as a vector in the same frame as the initial velocity $\xi'^1_0$, i.e. in the on-earth twin frame at the
initial crossing point 0. Then, the on-earth twin would consider his proper frame \textit{relatively\/} to the transmitted data and the telemetric (for
instance) evaluations as a Galilean frame, i.e. \textit{in abstractio\/} of his accelerations. But then, \textit{relatively\/} with respect to this inital frame, it will
correspond exactly to the true velocity vector $\xi'^1$ at $\tau-c\,\triangle(\tau)$. Then the integral \eqref{merged} can be computed at $\iota$ or at any
point on the worldline and the result compared with the observed time discrepancy (if $\tau'(\tau-c\,\triangle(\tau))$ is also transmitted and received at
$\tau$), giving the same result~: $\tau'-\tau+c\,\triangle(\tau)$.\par Hence we see all the complexity of such protocols.
\section{Conclusion and selected viewpoints}
\subsection{Determinism and Temperature}
Nevertheless, this computation would be known, before the experimental observation of the time discrepancy, only and only if $\xi'^1$ is a vector field and
not a set of vector values at $\xi$ of germs of vector fields on $\tom$, i.e. the system is deterministic, meaning that $\xi'^1$ is given by the flow map
$\Xi'$ such that $\xi'^1=\dot{\Xi}'(\eta'^1,s)$ where $\eta'^1$ is the value of $\xi'^1$ at $s=0$, i.e. at the origin $0$. On contrary, if
$\xi'^1$ is a value of germs which are not determined by the local geometry of
$\m$ (striated by the trajectory lines of the vector fields) but by external forces with an exogenous or endogenous origin, acting along the trajectory (we
think about a rocket producing external forces driven by a human being at will; Nevertheless, we could also consider ``programmable" routes defining
the paths, and then the system would remain deterministic in that particular case) the time discrepancy is undeterminable before the point
$\iota$.\par Moreover, we have to consider the paths in $T\tot$ because of the non-factorizable embeddings used to compute the time discrepancy, and we can
represent the lifted paths $\widehat{C}^1$ and  $\widehat{C}'^1$ as in the Figure 5 below which could be viewed as a Clapeyron cycle ! It would suggest a
link between the positivity of the entropy via the Clausius inequalities relative to the time discrepancy ascribing the $\dot{\xi}$'s as the so-called
\textit{``temperature 4-vectors"\/}.\par\medskip
Indeed,  from \eqref{normo} and \eqref{xiu} we deduce
\[
\|e^{a_\xi}\,\xi^1\|=
\|e^{a_{\xi'}}\,\xi'^1\|\,,
\]
where the norms $\|\dots\|$ are irrespectively taken from one of
the three metrics $h$, $h'$ or $h_0$. Then the vectors $u$ of the
``form" $u\equiv e^{a_\xi}\,\xi^1$ are invariant with respect to
``transported" Lorentz transformations, i.e. equivariant with
respect to diffeomorphism keeping equivariant the metrics; The
latter being consequently elements of the so-called Poincar{\'e}
pseudogroup. Obviously $u$ is the ``absolute" velocity vector
$\dot{\xi}$. Setting 
\[
T\equiv e^{a_\xi}
\] 
as the absolute
temperature at $\xi$, then we obtain the so-called
well-known ``temperature" 4-vector $u$ and, as a matter of
fact  and a  result, the parameter
$a_\xi$ is an entropy. Then it appears that falling in a potential
of acceleration of gravitation for instance, is as equivalent as an
increasing of entropy which then acquires the status usually
ascribed to interactions, i.e. an origin to a motion (or the
converse), and not only from a configuration. It also means somehow
that a 4D-ocean anisotropy is worked out from $a_\xi$ and the
entropy arrow is \textit{``an"\/} anisotropy non-geometrical (not in a 4
dimensional manifold) arrow in $\m$. Moreover at $p_0$, then
$T'=T$, i.e. the absolute temperature is a relativistic local
invariant. Also it follows that
\[
T_\xi f(u)=e^{(a_{\xi}-a_{\xi'})}u'\,,
\]
i.e. it is not a conformally equivariant vector, as well as the
absolute temperature.
\begin{center}
\fbox{\includegraphics[bb=16 161 415 725,scale=.4]{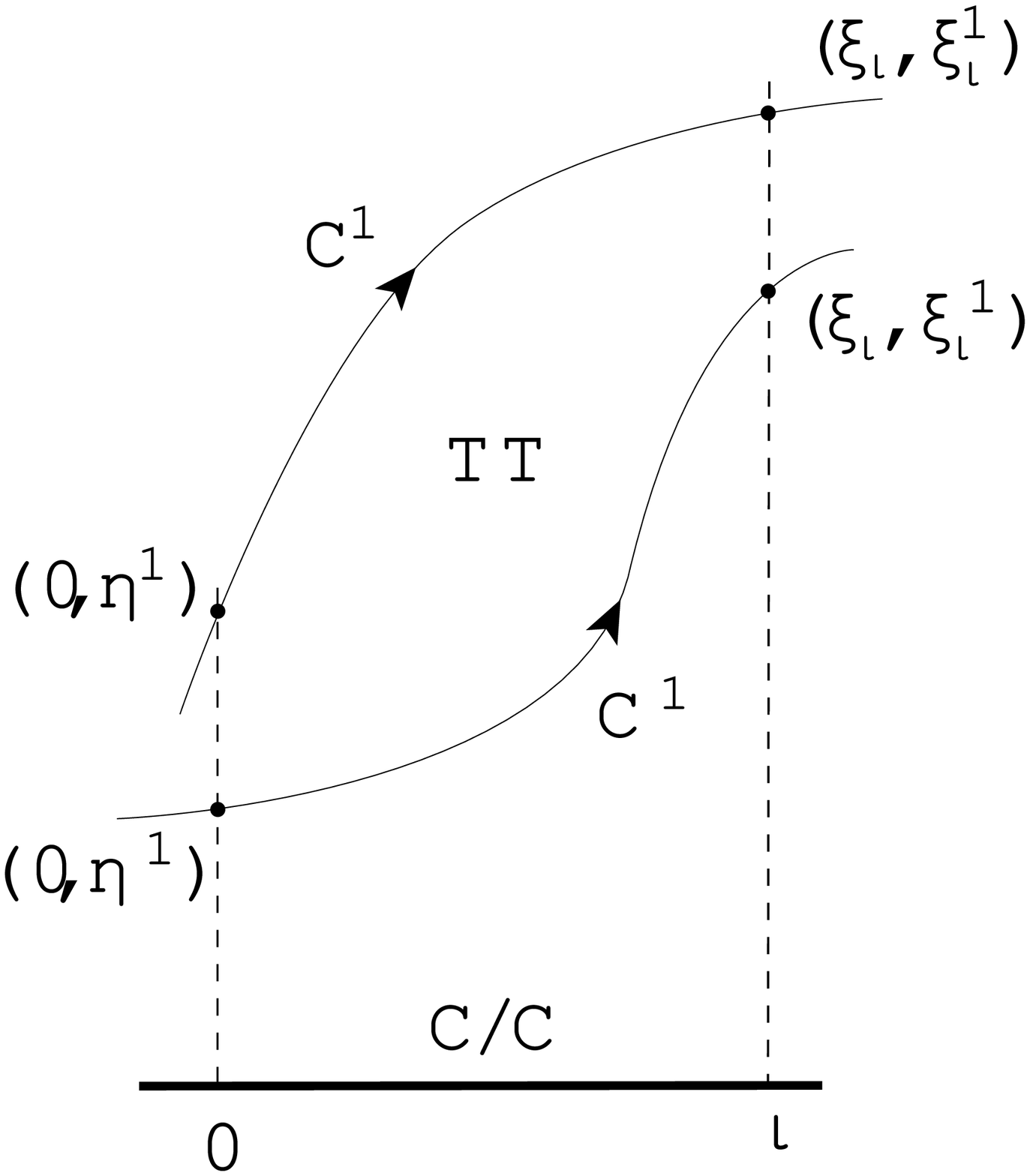}
Figure 5.
}
\end{center} Also the time
dilatation at 0 and $\iota$ due to a parallax effect, related to the Lorentz transformations $T_0f$ and $T_\iota f$, is kept, both at 0 and $\iota$. Indeed,
we have always a Lorentz transformation between two 4-vectors with same base point $p_0$, and these effects are completely included in the time discrepancy
formula \eqref{timetwintau}.
\subsection{Entropy}
The diagram above suggest a link with the thermodynamics. Then we wonder if entropy could be exhibited out of the mathematical treatment we made using
jets, germs and local rings of germs. As we demonstrated, it is rather impossible, in case of 0-jets $a_\xi$ or $a'_{\xi'}$ of functions, to compute as well
as predict consequently the time discrepancy $\triangle_{\iota}(\wl\,)$. Hence, we can defined a density of probability $p_\iota(\delta)$ for
$\triangle_{\iota}(\wl\,)$ to be equal to $\delta\in\Rset$ at the crossing ending point $\iota$. Then, it is easy to define the variation of entropy
$\triangle S$ occuring between
$0$ and $\iota$ by the usual formula coming from statistics ($k$ being the Boltzman constant)~:
\[
\triangle S=S_\iota-S_0\equiv -k\int_\Rset p_\iota(\delta)\,\ln(p_\iota(\delta))\,d\delta\,\geq0\,.
\]
\par\bigskip
 As a matter of fact, we didn't consider there exists a potential (a field of scalars)
$\alpha$ on $\tom$ such that $a_{\xi}=\alpha(\xi)$,
$a'_{\xi'}=\alpha(\xi')=\alpha\circ f(\xi)$, $a'_0=a_0$ and $a'_{\iota}=a_\iota$ which would define somehow 
``geodesic valleys" $\wc$ and $\wc'$ with 
moving particles or massive objects streaming through. Then $\triangle S$ would be vanishing if $\alpha\circ f(\xi)=\alpha(\xi)$ (i.e. we would have geodesic
trajectories with same starting and ending points in a gravitational field, and fields of 4-velocities), as well as the time discrepancy, since the density of
probability
$p_\iota(\delta)$ would be equal to 1 for $\delta=0$ only. But (!), we have to take care, in order to be rigourous also, that
a function $\alpha$ does not necessary physically  exist on $\tom$ or equivalently on $\m$ (i.e. as a
potential of acceleration), since it
is also determined by the travellers themselves. Indeed they can vary
their own acceleration as they wish and their own scalar
(not potential on $\m$) of acceleration ``$a$" by using rocket motors
for instance. It is a fundamental point which consists in
considering scalars fields, vectors fields or tensors fields not on $\m$ but on the paths only. In other
words, there are no mathematically pre-determined acceleration
scalars on $\m$ and consequently pre-determined manifold $\m$
(!) since the $a$'s can also be viewed as scalars of deformations of $\m$. 
This doesn't mean that determined potentials of
acceleration $\alpha$ don't exist in full generality. But they can for
instance, as a result of deterministic mass distributions. That is a kind of
direct consequence of the elevator metaphor defining the Einstein
equivalence principle in such a way that we could create by rocket
motors \dots a relative acceleration of gravity.\par It is a first source of
indeterminism, unless we consider the wishes of astronauts
satisfy some equations defining also what would be a
determined spacetime manifold
$\m$ (!). It would only be an idealized spacetime, and our brain would
only simulate a predicted spacetime from a memorized observed initial one,
satisfying the Einstein equations ! The masses distribution can't be determined
by the Einstein equations, and consequently the latter can't determine a
spacetime topology associated to a spacetime evolution or history via contingency. But then, it follows
that the geometry and the spacetime manifold is not the solution obtained from
the Einstein equation. Hence they can't bind a guenine spacetime geometry and a
masses distribution ! It could only be a simulated local ``static" spacetime
manifold, which have to be revised each time, time after time from new
experimental data. But the second is that no one (as far as we know\dots) is
ubiquous and henceforth can compute
$\triangle a$ because of a lack of datas (except in very
particular cases) being ``simultaneously" at $\xi$ and $\xi'$. 
\subsection{The light cones}
Also there is
an important problem: we  considered an unique metric
$h_0$ at $p_0$. This situation points out an other problem
in general relativity:  if
$\m$ is not foliated by a unique absolute time parameter, as we
assume, and which is equivalent to say that $\m$ is not endowed with
global metric fields, then two crossing paths such as in Figure 6
below, could be defined with two different light cones at the apex
$p_0$ in full generality. We avoid this situation in assuming that
$h$ and $h'$ are related to the unique metric $h_0$ in a unique way.
But the unicity of $h_0$ has to be explained as well as the
local spacetime anisotropy at $p_0$  it involves. 
\begin{center}
\noindent\fbox{\includegraphics[bb=155 384 407
671,scale=.7]{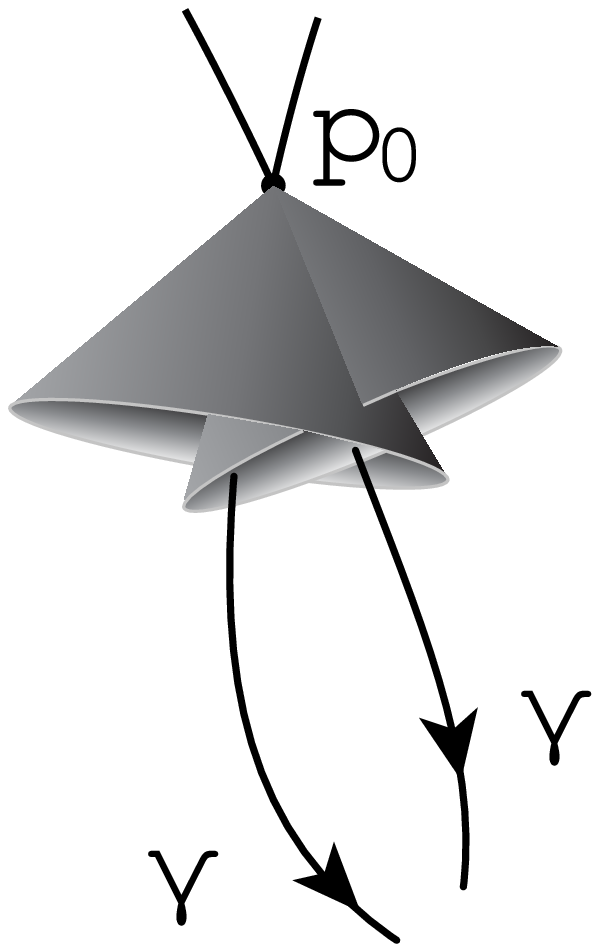} Figure 6.}
\end{center}\par\medskip
\section{Appendix}
In this appendix we recall the definitions of \textit{germs\/} and \textit{local rings\/} (see Atiyah, M. F. and Macdonald, I. G.~: \textit{Introduction to
Commutative Algebra\/}. Reading, MA: Addison-Wesley, 1969).\par Let $f\,,f'~: M\longrightarrow N$ be two differentiable maps of class $C^r$
($r\in\Nset\cup\{+\infty\}$ or $r=\omega$ for analytic maps), with $M$ and $N$ being differentiable manifolds of same class.
\begin{definition}
The two differentiable maps $f$ and $f'$ defined respectively in two neighborhoods $U$ and $U'$ of a same point $p\in M$, have the same \emph{``common germ at
$p$"\/} if and only if there are equal in a third neighborhood of $p$ (this third neighborhood can be included in the intersection $U\cap U'$). This a
relation of equivalence that we denote by $f\sim_p f'$.
\end{definition}
\begin{definition}
We call \emph{``germ"\/} of a differentiable map $f$, the class of differentiable maps $f'$ such that $f\sim_p f'$, and we use the notation $[f]_p$  for
the germ of $f$.
\end{definition}
Let $k$ be a field, and $V$ and $W$ two $k$-vector spaces. Moeover,
to each point $p$, element of a topological space $U$ (resp. $V$), is associated a ring (resp. a $k$-module) $\mathcal{O}_{U,p}$ (resp. $\mathcal{M}_{V,W,p}$)
of the germs of functions (resp. $k$-morphisms from $V$ to $W$) defined at $p$.
\begin{definition}
Let $\mathbb{M}_p$ be the ring (resp. $k$-module) of the non-invertible elements in  $\mathcal{O}_{U,p}$ (resp. $\mathcal{M}_{V,W,p}$).
If  $\mathbb{M}_p$ is the unique maximal ideal in $\mathcal{O}_{U,p}$ (resp. $\mathcal{M}_{V,W,p}$), then $\mathcal{O}_{U,p}$ (resp. $\mathcal{M}_{V,W,p}$)
is said to be a \textit{``local ring"\/} (\textit{``non-commutative local ring"\/}).
\end{definition}
We can outline these definitions with the very schematic drawing below~:
\begin{center}
\fbox{\includegraphics[scale=.5]{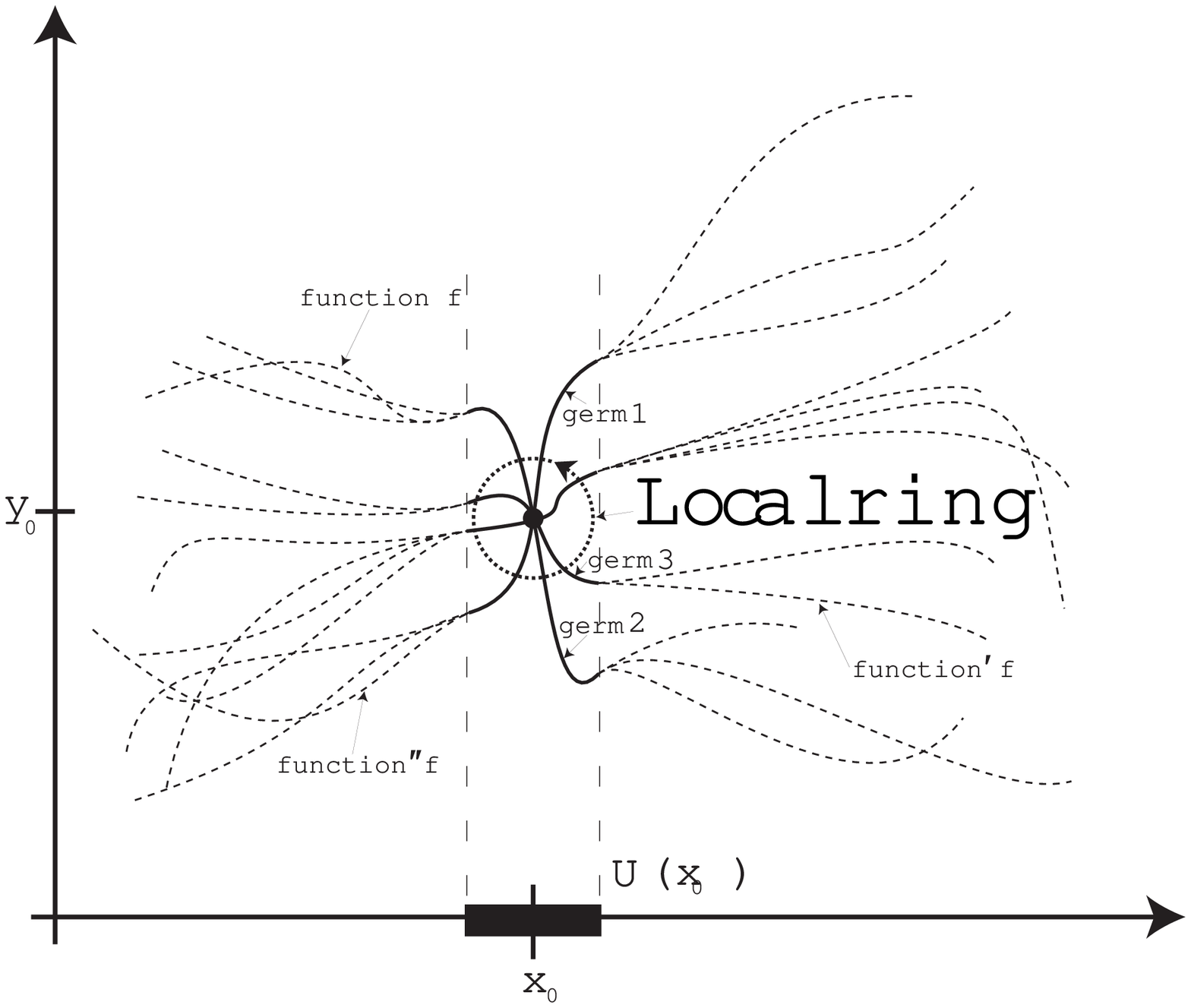}
Figure 7.}
\end{center}
\begin{definition}
Let $\nabla$ be a covariant derivative defined on an open set $U$ of a differentiable manifold $M$. We call \emph{``jet of order $k$"\/} or
\emph{``$k$-jet"\/} at a point
$x_0\in M$ of  a germ $[f]_{x_0}$ at $x_0$ of a differentiable application $f$ defined on an open neighbohood $U$ of $x_0$ , the full set of the $i$-th
covariant derivatives with respect to the vector $\eta$
\[\nabla^i_\eta[f]\equiv\underbrace{\nabla_\eta\circ\dots\circ\nabla_\eta}_{\mbox{$i$ \rm times}}\,[f]\]
of $[f]_{x_0}$ at $x_0$ with $0\leq i\leq k$. \emph{Usually, the derivatives are used in this definition instead of the covariant derivatives, but this can be
extended to the covariant derivatives case without modifications, but taking care that the covariant deivative could be not defined on the whole of $M$, but
at least on a neighbohood of $x_0$ only. Hence, we would have to consider a germ of covariant derivatives.\/}
\end{definition}
For instance, from the definition above, the value $f(x_0)$ is a 0-jet of a ``large" set of germs. In fact, it matters to notice that a given $k$-jet at $x_0$
of a germ of an application $f$ defines the \textit{local ring\/} of those diiferentiable applications with the same $k$-jet at $x_0$. Roughly speaking, a
$k$-jet at a given point could be viewed as  a set of germs, i.e. a local ring. And we consider this ascription in the whole of the present paper, since it
prohibits to consider values of maps or fields at a given point to be associated to a unique given (determined) map or field; The latter which would be
exhibited out of deterministic, \textit{ a priori\/}, given physical and/or geomerical global structures.
\bibliographystyle{alpha}
\bibliography{Twin_Paradox_Biblio}
\end{document}